\documentclass[12pt]{article}
\usepackage{graphicx}
\usepackage{lscape}
\usepackage{amssymb}
\usepackage{appendix}
\usepackage{multirow}
\usepackage{citesort}

\begin{document}
\def\qq{\langle \bar q q \rangle}
\def\uu{\langle \bar u u \rangle}
\def\dd{\langle \bar d d \rangle}
\def\sp{\langle \bar s s \rangle}
\def\GG{\langle g_s^2 G^2 \rangle}
\def\Tr{\mbox{Tr}}
\def\figt#1#2#3{
        \begin{figure}
        $\left. \right.$
        \vspace*{-2cm}
        \begin{center}
        \includegraphics[width=10cm]{#1}
        \end{center}
        \vspace*{-0.2cm}
        \caption{#3}
        \label{#2}
        \end{figure}
	}
	
\def\figb#1#2#3{
        \begin{figure}
        $\left. \right.$
        \vspace*{-1cm}
        \begin{center}
        \includegraphics[width=10cm]{#1}
        \end{center}
        \vspace*{-0.2cm}
        \caption{#3}
        \label{#2}
        \end{figure}
                }

\def\ds{\displaystyle}
\def\beq{\begin{equation}}
\def\eeq{\end{equation}}
\def\bea{\begin{eqnarray}}
\def\eea{\end{eqnarray}}
\def\beeq{\begin{eqnarray}}
\def\eeeq{\end{eqnarray}}
\def\ve{\vert}
\def\vel{\left|}
\def\ver{\right|}
\def\nnb{\nonumber}
\def\ga{\left(}
\def\dr{\right)}
\def\aga{\left\{}
\def\adr{\right\}}
\def\lla{\left<}
\def\rra{\right>}
\def\rar{\rightarrow}
\def\lrar{\leftrightarrow}  
\def\nnb{\nonumber}
\def\la{\langle}
\def\ra{\rangle}
\def\ba{\begin{array}}
\def\ea{\end{array}}
\def\tr{\mbox{Tr}}
\def\ssp{{\Sigma^{*+}}}
\def\sso{{\Sigma^{*0}}}
\def\ssm{{\Sigma^{*-}}}
\def\xis0{{\Xi^{*0}}}
\def\xism{{\Xi^{*-}}}
\def\qs{\la \bar s s \ra}
\def\qu{\la \bar u u \ra}
\def\qd{\la \bar d d \ra}
\def\qq{\la \bar q q \ra}
\def\gGgG{\la g^2 G^2 \ra}
\def\q{\gamma_5 \not\!q}
\def\x{\gamma_5 \not\!x}
\def\g5{\gamma_5}
\def\sb{S_Q^{cf}}
\def\sd{S_d^{be}}
\def\su{S_u^{ad}}
\def\sbp{{S}_Q^{'cf}}
\def\sdp{{S}_d^{'be}}
\def\sup{{S}_u^{'ad}}
\def\ssp{{S}_s^{'??}}

\def\sig{\sigma_{\mu \nu} \gamma_5 p^\mu q^\nu}
\def\fo{f_0(\frac{s_0}{M^2})}
\def\ffi{f_1(\frac{s_0}{M^2})}
\def\fii{f_2(\frac{s_0}{M^2})}
\def\O{{\cal O}}
\def\sl{{\Sigma^0 \Lambda}}
\def\es{\!\!\! &=& \!\!\!}
\def\ap{\!\!\! &\approx& \!\!\!}
\def\ar{&+& \!\!\!}
\def\ek{&-& \!\!\!}
\def\kek{\!\!\!&-& \!\!\!}
\def\cp{&\times& \!\!\!}
\def\se{\!\!\! &\simeq& \!\!\!}
\def\eqv{&\equiv& \!\!\!}
\def\kpm{&\pm& \!\!\!}
\def\kmp{&\mp& \!\!\!}
\def\mcdot{\!\cdot\!}
\def\erar{&\rightarrow&}


\def\simlt{\stackrel{<}{{}_\sim}}
\def\simgt{\stackrel{>}{{}_\sim}}


\renewcommand{\textfraction}{0.2}    
\renewcommand{\topfraction}{0.8}   

\renewcommand{\bottomfraction}{0.4}   
\renewcommand{\floatpagefraction}{0.8}
\newcommand\mysection{\setcounter{equation}{0}\section}

\def\baeq{\begin{appeq}}     \def\eaeq{\end{appeq}}  
\def\baeeq{\begin{appeeq}}   \def\eaeeq{\end{appeeq}}
\newenvironment{appeq}{\beq}{\eeq}   
\newenvironment{appeeq}{\beeq}{\eeeq}
\def\bAPP#1#2{
 \markright{APPENDIX #1}
\addcontentsline{toc}{section}{Appendix #1 #2}
 \medskip
 \medskip
\begin{center}      {\bf\LARGE Appendix #1 }{\quad\Large\bf #2}
\end{center}
 \renewcommand{\thesection}{#1.\arabic{section}}
\setcounter{equation}{0}
        \renewcommand{\thehran}{#1.\arabic{hran}}
\renewenvironment{appeq}
  {  \renewcommand{\theequation}{#1.\arabic{equation}}
     \beq
  }{\eeq}
\renewenvironment{appeeq}
  {  \renewcommand{\theequation}{#1.\arabic{equation}}
     \beeq
  }{\eeeq}
\nopagebreak \noindent}

\def\eAPP{\renewcommand{\thehran}{\thesection.\arabic{hran}}}

\renewcommand{\theequation}{\arabic{equation}}
\newcounter{hran}
\renewcommand{\thehran}{\thesection.\arabic{hran}}

\def\bmini{\setcounter{hran}{\value{equation}}
\refstepcounter{hran}\setcounter{equation}{0}
\renewcommand{\theequation}{\thehran\alph{equation}}\begin{eqnarray}}
\def\bminiG#1{\setcounter{hran}{\value{equation}}
\refstepcounter{hran}\setcounter{equation}{-1}
\renewcommand{\theequation}{\thehran\alph{equation}}
\refstepcounter{equation}\label{#1}\begin{eqnarray}}


\newskip\humongous \humongous=0pt plus 1000pt minus 1000pt
\def\caja{\mathsurround=0pt}


\title{
         {\Large
                 {\bf
Vertices of the heavy spin--3/2 sextet baryons with light vector mesons in QCD
                 }
         }
      }

\author{\vspace{1cm}\\
{\small T. M. Aliev$^1$ \thanks {e-mail:
taliev@metu.edu.tr}~\footnote{permanent address: Institute of
Physics, Baku, Azerbaijan}\,\,, K. Azizi$^2$ \thanks {e-mail:
kazizi@dogus.edu.tr}\,\,, M. Savc{\i}$^1$ \thanks
{e-mail: savci@metu.edu.tr}} \\
{\small $^1$ Physics Department, Middle East Technical University,
06531 Ankara, Turkey }\\
{\small$^2$ Physics Division,  Faculty of Arts and Sciences,
Do\u gu\c s University,} \\
{\small Ac{\i}badem-Kad{\i}k\"oy,  34722 Istanbul, Turkey}}

\date{}

\begin{titlepage}
\maketitle
\thispagestyle{empty}

\begin{abstract}
All transitions among the heavy spin--3/2 sextet baryons with participation of the
light vector mesons $(B_Q^* B_Q V)$ are investigated in the framework of the
light cone sum rules. These  vertices are described by four
coupling constants. The corresponding sum rules are derived for each coupling
constant. It is shown that the correlation functions of different transitions can be
described by only one invariant function.
\end{abstract}

\end{titlepage}

\section{Introduction}

During last few years, there have been significant progress in the
experimental studies of heavy baryons. Especially,  exciting discoveries have been made  in the heavy baryon
spectroscopy \cite{RBQsBQV01}.
The $\left({1\over 2}\right)^+$ antitriplet states 
$(\Lambda_c^+,~\Xi_c^+,~\Xi_c^0)$ as well as
the $\left({1\over 2}\right)^+$ and $\left({3\over 2}\right)^+$ sextet states
$(\Omega_c^0,~\Sigma_c,~\Sigma_c^\prime)$ and
$(\Omega_c^*,~\Sigma_c^*,~\Sigma_c^{*\prime})$ have been discovered, while
for the s--wave bottom baryons only the $\Lambda_b,~\Sigma_b,\Sigma_b^*$ and
$\Omega_b$ have been established \cite{RBQsBQV02}.

The experimental achievements simulate theoretical investigations in this
area. These baryons are very useful in understanding the dynamics of the
nonperturbative QCD. The heavy baryons provide rich laboratory for studying the
predictions of the heavy quark effective theory (HQET). The heavy baryons are also very
informative in investigating the polarization effects. This is due to the
fact that part of the polarization of heavy quark is transferred to the
heavy baryon.

Following the experimental discoveries of heavy hadrons, the interest of researchers
 is focused on the study of their weak, electromagnetic and strong decays.    
The semileptonic weak decays of heavy baryons can provide us invaluable
knowledge about the Cabibbo--Kobayashi--Maskawa (CKM) matrix elements and
information about their internal structures. In studying the strong decays
of heavy baryons, one needs to know the corresponding strong coupling constants. As is well
known, formation of hadrons take place at low energy scale
belongs to the nonperturbative region of QCD. Therefore, we can not calculate
the strong coupling constants of baryons starting from the fundamental
Lagrangian. For this reason, in calculating the strong coupling constants,
some nonperturbative approachs are needed. The QCD sum rules method
\cite{RBQsBQV03} is one of the most predictive one among a bunch of
nonperturbative approaches. In the present work, we calculate the strong
coupling constants of the transitions among sextet spin--3/2 baryons within
the QCD light cone sum rules (LCSR) \cite{RBQsBQV04}. 
The main difference of the light cone and traditional versions of the sum rules
is that, in the LCSR the operator product expansion (OPE) is performed over
the twists rather than dimension of the operators, as is the case in the
latter one. It should be noted that the coupling constants of heavy
spin--1/2 baryons with light vector and pseudoscalar mesons are calculated
within the LCSR method in \cite{RBQsBQV05,RBQsBQV06}, the coupling 
constants of spin--3/2 to spin--1/2 heavy baryon--light mesons are
calculated in \cite{RBQsBQV07,RBQsBQV08}, and the coupling constants of
spin--3/2 sextet heavy baryons with pseudoscalar mesons are calculated in
\cite{RBQsBQV09}.

The present study is organized in the following way. 
In section 2,   first we obtain the structure independent relations among the
correlation functions. Then, we derive  sum rules for the coupling
constants of the spin--3/2 sextet baryons with light vector
mesons in terms of the invariant functions. In section 3, the numerical
analysis of the sum rules obtained in section 2 is carried out.
   
\section{Sum rules for the $B_Q^* B_Q^* V$ couplings in LCSR}

This section is devoted to the calculation of the strong coupling constants
among the heavy spin--3/2 sextet baryons with light vector mesons. For
determination of these coupling constants within LCSR, the following
correlation function is introduced:
\bea
\label{eBQsBQV01}
\Pi_{\mu\nu} = i \int d^4x e^{ipx} \lla V(q) \vel {\cal T} \Big\{
\eta_\mu (x) \bar \eta_\nu (0) \Big\} \ver 0 \rra~,
\eea
where $V(q)$ is the light vector meson with four momentum $q$, and
$\eta_\mu$ and $\eta_\nu$ are the interpolating currents for the sextet
heavy spin--3/2 baryons. The explicit form of interpolating current for the
heavy spin--3/2 baryons is given as \cite{RBQsBQV10}
\bea
\label{eBQsBQV02} 
\eta_\mu = A \epsilon^{abc} \Big\{ \ga q_1^{aT} C \gamma_\mu q_2^b \dr Q^c
+ \ga q_2^{aT} C \gamma_\mu Q^b \dr q_1^c +
\ga Q^{aT} C \gamma_\mu q_1^b \dr q_2^c \Big\}~.
\eea
where the normalization constant $A$ and the light quark content of all members are given in Table 1.
\begin{table}[thb]

\renewcommand{\arraystretch}{1.3}
\addtolength{\arraycolsep}{-0.5pt}
\small
$$
\begin{array}{|l|c|c|c|c|c|c|}
\hline \hline
 & \Sigma_{b(c)}^{*+(++)} & \Sigma_{b(c)}^{*0(+)} & \Sigma_{b(c)}^{*-(0)}
 & \Xi_{b(c)}^{*0(+)}    & \Xi_{b(c)}^{*-(0)}
 & \Omega_{b(c)}^{*-(0)}          \\  \hline
 q_1 & u & u & d & u & d & s \\
 q_2 & u & d & d & s & s & s  \\
 A   & \sqrt{1/3} & \sqrt{2/3} & \sqrt{1/3}
     & \sqrt{2/3} & \sqrt{2/3} & \sqrt{1/3} \\
\hline \hline
\end{array}
$$
\caption{The normalization constant and the light quark content of the 
heavy spin--3/2 sextet baryons.}
\renewcommand{\arraystretch}{1}
\addtolength{\arraycolsep}{-1.0pt}
\end{table}

According to the QCD sum rules philosophy, this correlation function is
calculated in terms of hadrons from one side (phenomenological) and in terms
of quark--gluon degrees of freedom from the other side (theoretical). In
order to obtain the phenomenological side of the sum rules, the correlation
function is saturated by a complete set of hadronic states that carries the
same quantum numbers as the interpolating current $\eta_\mu$. Isolating the
ground state contribution one can easily obtain
\bea
\label{eBQsBQV03}
\Pi_{\mu\nu} =  {\lla 0 \vel \eta_{\mu}(0) \ver
B_Q^*(p_2) \rra \lla B_Q^*(p_2) V(q) \vel \right.
B_Q^*(p_1) \rra \lla B_Q^*(p_1) \vel \bar{\eta}_{\nu}(0) \ver
0 \rra \over \ga p_2^2-m_2^2 \dr \ga p_1^2-m_1^2\dr}
+ \cdots~,
\eea
In all following discussion we put $p_2=p$ and $p_1=p+q$. In order to obtain
the expression of the correlation function from phenomenological side, we need
to know the matrix elements entering into Eq. (\ref{eBQsBQV03}). These
matrix elements are defined as
\bea
\label{eBQsBQV04}
\lla 0 \vel \eta_{\mu}\ver B_Q^*(p+q) \rra \es \lambda_{B_Q^*} u_{\mu}(p) ~,
\eea
where $\lambda_{B_Q^*}$ is the residue of $B_Q^*$ heavy baryon and $u_\mu(p)$
is the Rarita--Schwinger spinor. The matrix element $\lla B_Q^*(p) V(q)
\vel \right. B_Q^*(p+q) \rra$ is determined with the help of four form
factors $g_1$, $g_2$, $g_3$ and $g_4$ in the following way:
\bea
\label{eBQsBQV05}
\lla B_Q^*(p) V(q) \vel \right. B_Q^*(p+q) \rra \es \bar{u}_\alpha (p)
\Bigg\{ g^{\alpha\beta} \Bigg[ \rlap/\varepsilon g_1 + 2 p \varepsilon {g_2
\over m_1+m_2} \Bigg] \nnb \\
\ar {q^\alpha q^\beta \over (m_1+m_2)^2} \Bigg[\rlap/\varepsilon g_3 + 2 p
\varepsilon {g_4\over m_1+m_2} \Bigg] \Bigg\} u_\beta (p+q)~,
\eea
where $\varepsilon_\mu$ is the four polarization of the light vector meson.

Substituting Eqs. (\ref{eBQsBQV04}) and (\ref{eBQsBQV05}) into Eq.
(\ref{eBQsBQV03}), and performing summation over spins of spin--3/2 baryons
with the help of the formula
\bea
\label{eBQsBQV06}
\sum_s u_\mu (p,s) \bar{u}_\nu (p,s) = ( {\rlap/p +m } )\Bigg(
g_{\mu\nu} - {1\over 3} \gamma_\mu \gamma_\nu + {2 p_\mu p_\nu \over 3
m^2} + {p_\mu \gamma_\nu - p_\nu \gamma_\mu \over 3 m} \Bigg)~,
\eea
we can obtain the physical side of the correlation function.

At this point we face two principal drawbacks: a) the structures that appear
are not all independent, b) the interpolating current for the heavy spin--3/2 baryon  
couples to the spin--1/2 states, i.e.,
\bea                                                                        
\label{eBQsBQV07}
\lla 0 \vel \eta_{\mu}\ver 1/2(p) \rra \es A \ga \gamma_\mu - {4\over m}
p_\mu \dr u(p)~.
\eea
It follows from this equation that the structures that proportional to
$\gamma_\mu$ on the left and $\gamma_\nu$ on the right, as well as the terms
that are proportional to $p_\mu$ and $(p+q)_\nu$ contain contributions
coming from unwanted spin--1/2 states, and hence they should be removed.
Both of the above--mentioned problems can be solved by ordering the Dirac
matrices. In the present work, we choose the ordering of the Dirac matrices
in the form $\gamma_\mu \rlap/\varepsilon \rlap/q \rlap/p \gamma_\nu$.

Taking into account this procedure, we get the following expression for the
physical side of the correlation function:
\bea
\label{eBQsBQV08}
\Pi_{\mu\nu} \es { \lambda_{B_{Q_1}^*} \lambda_{B_{Q_2}^*} \over
[(p+q)^2-m_1^2)] (p^2 -  m_2^2)} \Bigg\{
2 (\varepsilon p) g_{\mu\nu} \rlap/q \Bigg[ g_1 + 
g_2 {m_2\over m_1+m_2}\Bigg] \nnb \\
\ek 2 (\varepsilon p) g_{\mu\nu} \rlap/q \rlap/p {g_2\over
m_1+m_2} + q_\mu q_\nu \rlap/\varepsilon \rlap/q \rlap/p 
{g_3 \over (m_1+m_2)^2} -
2 (\varepsilon p) q_\mu q_\nu \rlap/\varepsilon \rlap/q \rlap/p {g_4 \over
(m_1+m_2)^3} \nnb \\
\ar \mbox{\rm other structures beginning with $\gamma_\mu$ and ending with
$\gamma_\nu$, or} \nnb \\
& &\mbox{\rm terms that are proportional to $p_\mu$ or $(p+q)_\nu$} \Bigg\}~.
\eea

In order to obtain sum rules for the  combination $g_1+g_2{m_2\over m_1+m_2}$ and form factors   $g_2$, $g_3$
and $g_4$, we choose the coefficients of the structures, $(\varepsilon p)
g_{\mu\nu} \rlap/q$, $(\varepsilon p) g_{\mu\nu} \rlap/q \rlap/p$,
$q_\mu q_\nu \rlap/\varepsilon \rlap/q \rlap/p$ and $(\varepsilon p) q_\mu
q_\nu \rlap/\varepsilon \rlap/q \rlap/p$, respectively.

As has already been noted, in order to obtain the sum rules for the coupling
constants, the correlation function from the QCD side is needed. Before
calculating the theoretical part of the correlation function we obtain the
relations among invariant functions, which are quite efficient in calculation
of the coupling constants $g_i$. For this purpose, we shall follow  the
works \cite{RBQsBQV05,RBQsBQV06,RBQsBQV07,RBQsBQV08,RBQsBQV09},
where essential points of the relevant approach are
presented. The main advantage of this approach is that it involves $SU(3)_f$
symmetry violation effects, as well as the fact that the obtained relations among invariant functions
are all structure independent.

As an example, we consider the transition $\Sigma_b^{*0} \rar \Sigma_b^{*0}
\rho^0$. The invariant function corresponding to any structure can formally
be written in the following form:
\bea
\label{eBQsBQV09}
\Pi^{\Sigma_b^{*0} \rar \Sigma_b^{*0} \rho^0} = g_{\rho^0\bar{u}u} \Pi_1(u,d,b) +
g_{\rho^0\bar{d}d} \Pi_1^{'}(u,d,b) + g_{\rho^0\bar{b}b} \Pi_2(u,d,b)~.
\eea
It follows from Eq. (\ref{eBQsBQV02}) that the interpolating currents for the
heavy spin--3/2 baryons are symmetric with respect to the light quark
interchange, and therefore $\Pi_1^{'}(u,d,b) = \Pi_1(d,u,b)$. Using the form
of the interpolating current of $\rho^0$ meson, one can easily see that
\bea
\label{nolabel}
g_{\rho^0 \bar{u}u} \es - g_{\rho^0 \bar{d}d} = {1\over \sqrt{2}}~, g_{\rho^0 \bar{b}b} = 0~,
\eea
as a result of which we get,
\bea
\label{eBQsBQV10}
\Pi^{\Sigma_b^{*0} \rar \Sigma_b^{*0} \rho^0} = {1\over \sqrt{2}} \Big[
\Pi_1(u,d,b) - \Pi_1(d,u,b) \Big]~.
\eea
Obviously, $\Pi^{\Sigma_b^{*0} \rar \Sigma_b^{*0} \rho^0} = 0$ in the 
$SU(2)_f$ limit.

Taking into account the quark content of $\omega$ and $\phi$ mesons we
observe that
\bea
\label{nolabel}
g_{\omega \bar uu} \es - g_{\omega \bar dd} = {1\over \sqrt{2}}~, 
g_{\phi \bar ss} = 1~.
\eea
The functions $\Pi_1$, $\Pi_1^{'}$ and $\Pi_2$  in Eq. (\ref{eBQsBQV09}) physically
correspond to radiation of $\rho$ meson from $u,~d$ and $b$ quarks,
respectively.

 Following the works 
\cite{RBQsBQV05,RBQsBQV06,RBQsBQV07,RBQsBQV08,RBQsBQV09},
one can obtain
relations among the invariant functions involving $\rho,~\omega,~K^*$ and
$\phi$ mesons. These relations are given in the Appendix.
It follows from the relations among the invariant functions presented in the
main body of the text and in the appendix that all these transitions can be
described in terms of a single universal function.

Now we turn back to our main problem, i.e., constructing the sum rules for
the strong coupling constants describing the transitions among heavy
spin--3/2 sextet baryons. Using the expressions of distribution
amplitudes (DA's) for the light vector 
mesons, as well as the quark operators, the theoretical part of the sum rules
can, in principle, be obtained in the standard way.

The theoretical part of the correlation function can be calculated in deep
Eucledian region, $-p^2 \rar \infty$, $-(p+q)^2 \rar \infty$, using the OPE.
The main nonperturbative inputs of the LCSR method is the  DA's. In the problem under consideration, we need to know the
DA's of the light vector mesons, which are given in
\cite{RBQsBQV11,RBQsBQV12,RBQsBQV13}. In order to calculate the theoretical
part of the correlation function, the expressions of the heavy and light
quark propagators are needed, whose expressions are given in
\cite{RBQsBQV14} and \cite{RBQsBQV15}, respectively. 

The final step for obtaining sum rules of the strong coupling constants, is
equating the coefficients of the structures 
$(\varepsilon p) g_{\mu\nu} \rlap/q$, 
$(\varepsilon p) g_{\mu\nu} \rlap/q \rlap/p$,
$q_\mu q_\nu \rlap/{\varepsilon} \rlap/q \rlap/p$ and
$(\varepsilon p) q_\mu q_\nu \rlap/q \rlap/p$ from both representations of
the correlation function, and then applying double Borel transformation on
the variables $-p^2$ and $-(p+q)^2$ on both sides, which suppresses the
contributions of the higher states and continuum. As a result of these
operations, we obtain the following sum rules for the strong coupling
constants $g_i$:

\bea
\label{eBQsBQV11}
g_1+g_2{m_2  \over m_1+m_2} \es {1\over 2 \lambda_{B_{Q_1}^*} \lambda_{B_{Q_2}^*}} 
e^{ {m_1^2\over M_1^2} + {m_2^2\over M_2^2} + {m_V^2\over M_1^2+M_2^2} }
\Pi_1^{(1)}~, \nnb \\
g_2 \es - {m_1+m_2 \over 2 \lambda_{B_{Q_1}^*} \lambda_{B_{Q_2}^*}}    
e^{ {m_1^2\over M_1^2} + {m_2^2\over M_2^2} + {m_V^2\over M_1^2+M_2^2} }
\Pi_1^{(2)}~, \nnb \\
g_3 \es - {(m_1+m_2)^2 \over \lambda_{B_{Q_1}^*} \lambda_{B_{Q_2}^*}}  
e^{ {m_1^2\over M_1^2} + {m_2^2\over M_2^2} + {m_V^2\over M_1^2+M_2^2} }
\Pi_1^{(3)}~, \nnb \\
g_4 \es - {(m_1+m_2)^3 \over 2 \lambda_{B_{Q_1}^*} \lambda_{B_{Q_2}^*}}                     
e^{ {m_1^2\over M_1^2} + {m_2^2\over M_2^2} + {m_V^2\over M_1^2+M_2^2} }
\Pi_1^{(4)}~,
\eea
where $M_1^2$ and $M_2^2$ are the Borel parameters in initial and final
channels, respectively. It should be remembered that the relations among
the invariant functions are structure independent, but their explicit
expressions are structure dependent. For this reason, we provide the
invariant functions with one extra upper index. The indicies  1, 2, 3 and 4 correspond
to the choice of the structures $(\varepsilon p) g_{\mu\nu} \rlap/q$,
$(\varepsilon p) g_{\mu\nu} \rlap/q \rlap/p$, 
$q_\mu q_\nu \rlap/\varepsilon \rlap/q \rlap/p$ and $(\varepsilon p) q_\mu
q_\nu \rlap/\varepsilon \rlap/q \rlap/p$, respectively. In further numerical
analysis, we set $M_1^2=M_2^2=M^2$ since the masses of the initial and final
baryons are very close to each other. The residues ofthe  heavy spin--3/2 baryons
have been estimated within the QCD sum rules in \cite{RBQsBQV16}.  

\section{Numerical analysis}

This section is devoted to the numerical calculation of the sum rules for
the coupling constants of the spin--3/2 to spin--3/2 heavy baryon
transitions with the participation of the light vector mesons.
The main input parameters entering  the sum rules are the Borel mass
parameter $M^2$, the continuum threshold $s_0$ and DA's of the light vector
mesons. The expresions of DA's  are taken
from \cite{RBQsBQV11,RBQsBQV12,RBQsBQV13,RBQsBQV14}.  The values of
other input parameters are: $\lla 0 \vel {\alpha_s\over \pi} G^2 \ver 0
\rra = (0.012 \pm 0.01)~GeV^4$, $\lla \bar{u}u \rra = \lla \bar{d}d \rra =
- (0.24 \pm 0.01)^3~GeV^3$, $\lla \bar{s}s \rra = 0.8 \lla \bar{u}u \rra$
\cite{RBQsBQV17}, $m_0^2 = (0.8\pm0.2)~GeV^2$ \cite{RBQsBQV18},
 $m_s(2~GeV)=(111 \pm 6)~MeV$ at $\Lambda_{QCD} = 330~MeV$
\cite{RBQsBQV19}. For the masses  of the heavy hadrons
we use the results of the work \cite{RBQsBQV20}.

The continuum threshold $s_0$ and the Borel mass $M^2$ are the auxiliary
parameters of the considered sum rules. For this reason, we should find the
so--called "working region'' of these parameters, where physical quantities
are practically independent of them. The working region of $M^2$ is
determined in the following way. The upper limit of $M^2$ is obtained by
requiring that the contributions of higher states and continuum should be
less than 40\% of the total result of the correlation function. The lower
limit of $M^2$ is determined by demanding that the contribution of the
terms with higher powers of $1/M^2$ constitute (20--25)\% of the contributions
from the terms with highest power of $M^2$. Taking both these
conditions into account, we find that the "working region'' of $M^2$ for the baryons with
$b$--quark lies in the region $10 \le M^2 \le 20~GeV^2$, while for the baryons
containing $c$--quark it is $4 \le M^2 \le 8~GeV^2$. The continuum threshold
$s_0$  depends on the mass of the first excited state. In general, in the quark models,
the energy difference between the first excited and ground states is about
$0.5~GeV$. Therefore, for the continuum threshold we use $s_0\approx
(m_{ground} + 0.5)^2~GeV^2$. This leads us to choose the ''working region" of the
continuum threshold as $38~GeV^2 \le s_0 \le 42~GeV^2$ and $9~GeV^2 \le s_0
\le 12~GeV^2$ for the heavy baryons with $b$ and $c$ quarks, respectively.

In the present work, we study the dependence of the coupling constants $g_1$,
$g_2$, $g_3$ and $g_4$ on the Borel mass parameter $M^2$ in the range
determined by its own working region, at several different values of $s_0$.   
As an example, in Figs. (1)--(4) we present the results of our numerical
analysis for the $\Xi_b^{*0} \rar \Xi_b^{*0} \rho^0$ transition. We see from
these figures that the coupling constants exhibit good stability on $M^2$
and  $s_0$.
The predictions of the sum rules for the coupling constants $g_1$,
$g_2$, $g_3$ and $g_4$ are presented in Tables (1) and (2). Note that in the
Tables, we only present the modules of the strong coupling constants, since the sum
rules method cannot predict the sign of the residues of the heavy baryons.

\begin{table}[thb]

\renewcommand{\arraystretch}{1.3}
\addtolength{\arraycolsep}{-0.5pt}
\small
$$
\begin{array}{|l|r@{\pm}l|r@{\pm}l|r@{\pm}l|r@{\pm}l|}
\hline \hline  
\mbox{\small{~~~~\, transition}}       &
     \multicolumn{2}{c|}{\mbox{$g_1$}}  &
     \multicolumn{2}{c|}{\mbox{$g_2$}}  &
     \multicolumn{2}{c|}{\mbox{$g_3$}}  &
     \multicolumn{2}{c|}{\mbox{$g_4$}}   \\ \hline
\Xi^{\ast 0}_b      \rar  \Xi^{\ast 0}_b      \rho^0              &  9  &  1  & 19 & 2  & 50 &  5  & 15 &  2   \\
\Sigma^{\ast 0}_b   \rar  \Sigma^{\ast -}_b   \rho^+              & 18  &  2  & 36 & 4  &100 & 20  & 27 &  3   \\ 
\Xi^{\ast 0}_b      \rar  \Sigma^{\ast +}_b    K^{\ast -}         & 30  &  5  & 40 & 4  &110 & 15  & 28 &  3   \\
\Omega^{\ast -}_b   \rar  \Xi^{\ast 0}_b       K^{\ast -}         & 30  &  5  & 41 & 4  & 110 & 20  & 28 &  3   \\
\Sigma^{\ast +}_b   \rar  \Xi^{\ast 0}_b       K^{\ast +}         & 30  &  5  & 40 & 4  &100 & 20  & 28 &  3   \\
\Xi^{\ast 0}_b      \rar  \Omega^{\ast -}_b    K^{\ast +}         & 35  &  5  & 42 & 5  &100 & 20  & 28 &  3   \\
\Sigma^{\ast +}_b   \rar  \Sigma^{\ast +}_b   \omega              & 16  &  2  & 32 & 3  & 90 & 10  & 24 &  4   \\ 
\Xi^{\ast 0}_b      \rar  \Xi^{\ast 0}_b      \omega              & 10  &  2  & 17 & 2  & 45 &  5  & 13 &  2   \\
\Xi^{\ast 0}_b      \rar  \Xi^{\ast 0}_b      \phi                & 17  &  2  & 25 & 3  & 85 & 10  & 25 &  5   \\
\Omega^{\ast -}_b   \rar  \Omega^{\ast -}_b   \phi                & 40  &  5  & 50 & 6  &140 & 20  & 50 & 10   \\
 \hline \hline
\end{array}
$$
\caption{Coupling constants of the light vector mesons with heavy spin--3/2
baryons containing $b$ quark.}
\renewcommand{\arraystretch}{1}
\addtolength{\arraycolsep}{-1.0pt}

\end{table}


\begin{table}[thb]

\renewcommand{\arraystretch}{1.3}
\addtolength{\arraycolsep}{-0.5pt}
\small
$$
\begin{array}{|l|r@{\pm}l|r@{\pm}l|r@{\pm}l|r@{\pm}l|}
\hline \hline  
\mbox{\small{~~~~\, transition}}       &
     \multicolumn{2}{c|}{\mbox{$g_1$}}  &
     \multicolumn{2}{c|}{\mbox{$g_2$}}  &
     \multicolumn{2}{c|}{\mbox{$g_3$}}  &
     \multicolumn{2}{c|}{\mbox{$g_4$}}   \\ \hline
\Xi^{\ast +}_c      \rar  \Xi^{\ast +}_c      \rho^0              &  8  &  1  & 13 & 1  & 27 &  3  & 16 & 2   \\
\Sigma^{\ast +}_c   \rar  \Sigma^{\ast 0}_c   \rho^+              & 15  &  1  & 18 & 1  & 45 &  5  & 10 & 2   \\ 
\Xi^{\ast +}_c      \rar  \Sigma^{\ast ++}_c   K^{\ast -}         & 29  &  2  & 19 & 1  & 50 & 10  & 13 & 2   \\
\Omega^{\ast 0}_c   \rar  \Xi^{\ast +}_c       K^{\ast -}         & 18  &  2  & 26 & 4  & 50 & 10  & 30 & 4   \\
\Sigma^{\ast ++}_c  \rar  \Xi^{\ast +}_c       K^{\ast +}         & 27  &  4  & 19 & 1  & 50 & 10  & 13 & 2   \\
\Xi^{\ast +}_c      \rar  \Omega^{\ast 0}_c    K^{\ast +}         & 21  &  6  & 27 & 2  & 52 & 10  & 31 & 4   \\
\Sigma^{\ast ++}_c  \rar  \Sigma^{\ast ++}_c  \omega              & 13  &  2  & 16 & 2  & 40 &  5  &  9 & 2   \\ 
\Xi^{\ast +}_c      \rar  \Xi^{\ast +}_c      \omega              & 10  &  2  & 11 & 2  & 23 &  3  & 17 & 3   \\
\Xi^{\ast +}_c      \rar  \Xi^{\ast +}_c      \phi                & 13  &  2  & 12 & 1  & 38 &  5  &  9 & 2   \\
\Omega^{\ast 0}_c   \rar  \Omega^{\ast 0}_c   \phi                & 36  &  6  & 35 & 5  & 80 & 10  & 45 & 5   \\
 \hline \hline
\end{array}
$$
\caption{The same as Table (1), but for the heavy baryons containing $c$
quark.}
\renewcommand{\arraystretch}{1}
\addtolength{\arraycolsep}{-1.0pt}

\end{table}


The errors given in the  Tables (1) and (2) can be attributed to the
uncertainties of the input parameters and uncertainties inherit in
$M^2$ and $s_0$. At the end of
this section, it should be noted  that part of the relevant coupling
constants are calculated within the same framework in \cite{RBQsBQV21},
which are considerably different compared to the results presented in this
work. In our opinion, these discrepancies between our results and those that
are given in \cite{RBQsBQV21} could be due to the following reasons: the
residues of $\Omega_Q^*$, $\Xi_Q^*$, $\Sigma_Q^*$ given in \cite{RBQsBQV21}
are several times of magnitude larger compared to ours. Moreover, the
results presented in \cite{RBQsBQV21} do not satisfy the relations between
invariant functions. 

In conclusion,  in the present work, we have calculated the coupling
constants of heavy sextet spin--3/2 to spin--3/2 transitions with the
participation of the light vector mesons in LCSR. It is shown that the
relations among the correlation functions responsible for  different
transitions are described in terms of only one universal function. These
relations are all structure independent, while the explicit expression of
this universal function is structure dependent. The values of the four
coupling constants that appear in the parametrization of the $B_Q^* B_Q^* V$
vertex are obtained.


\newpage

\bAPP{}{}

In this appendix we present the expressions of the correlation functions
in terms of invariant function $\Pi_1$.

\baeeq
\label{nolabel}
\Pi^{\Sigma_b^{\ast +} \rar \Sigma_b^{\ast +}     \rho^0 } \es
\sqrt{2} \Pi_1(u,u,b)~, \nnb \\
\Pi^{\Sigma_b^{\ast -} \rar \Sigma_b^{\ast -}     \rho^0 } \es
- \sqrt{2} \Pi_1(d,d,b)~, \nnb \\
\Pi^{\Xi_b^{\ast 0} \rar \Xi_b^{\ast 0}     \rho^0 } \es 
{1\over \sqrt{2}} \Pi_1(u,s,b)~, \nnb \\
\Pi^{\Xi_b^{\ast -} \rar \Xi_b^{\ast -}     \rho^0 } \es 
- {1\over \sqrt{2}} \Pi_1(d,s,b)~, \nnb \\
\Pi^{\Sigma_b^{\ast +} \rar \Sigma_b^{\ast 0}     \rho^+ } \es 
\sqrt{2} \Pi_1(d,u,b)~, \nnb \\
\Pi^{\Sigma_b^{\ast 0} \rar \Sigma_b^{\ast -}     \rho^+ } \es 
\sqrt{2} \Pi_1(u,d,b)~, \nnb \\
\Pi^{\Xi_b^{\ast 0} \rar \Xi_b^{\ast -}     \rho^+ } \es 
\Pi_1(d,s,b)~, \nnb \\
\Pi^{\Sigma_b^{\ast 0} \rar \Sigma_b^{\ast +}     \rho^- } \es
\sqrt{2} \Pi_1(d,u,b)~, \nnb \\
\Pi^{\Sigma_b^{\ast -} \rar \Sigma_b^{\ast 0}     \rho^- } \es 
\sqrt{2} \Pi_1(u,d,b)~, \nnb \\
\Pi^{\Xi_b^{\ast -} \rar \Xi_b^{\ast 0}     \rho^- } \es 
\Pi_1(u,s,b)~, \nnb \\
\Pi^{\Xi_b^{\ast 0} \rar \Sigma_b^{\ast +}     K^{\ast-}} \es 
\sqrt{2} \Pi_1(u,u,b)~, \nnb \\
\Pi^{\Xi_b^{\ast -} \rar \Sigma_b^{\ast 0}     K^{\ast-}} \es 
\Pi_1(u,d,b)~, \nnb \\
\Pi^{\Omega_b^{\ast -} \rar \Xi_b^{\ast 0}     K^{\ast-}} \es 
\sqrt{2} \Pi_1(s,s,b)~, \nnb \\
\Pi^{\Sigma_b^{\ast +} \rar \Xi_b^{\ast 0}     K^{\ast+}} \es 
\sqrt{2} \Pi_1(u,u,b)~, \nnb \\
\Pi^{\Sigma_b^{\ast 0} \rar \Xi_b^{\ast -}     K^{\ast+}} \es 
\Pi_1(u,d,b)~, \nnb \\
\Pi^{\Xi_b^{\ast 0} \rar \Omega_b^{\ast -}      K^{\ast+}} \es 
\sqrt{2}\Pi_1(s,s,b)~, \nnb \\
\Pi^{\Xi_b^{\ast 0} \rar \Sigma_b^{\ast 0}     \bar{K}^{\ast0}} \es 
\Pi_1(d,u,b)~, \nnb \\
\Pi^{\Xi_b^{\ast -} \rar \Sigma_b^{\ast -}     \bar{K}^{\ast0}} \es 
\sqrt{2} \Pi_1(d,d,b)~, \nnb \\
\Pi^{\Omega_b^{\ast -}  \rar \Xi_b^{\ast -}    \bar{K}^{\ast0}} \es 
\sqrt{2} \Pi_1(s,s,b)~, \nnb \\
\Pi^{\Sigma_b^{\ast 0} \rar \Xi_b^{\ast 0}     K^{\ast0}} \es 
\Pi_1(d,u,b)~, \nnb \\
\Pi^{\Sigma_b^{\ast -} \rar \Xi_b^{\ast -}     K^{\ast0}} \es 
\sqrt{2} \Pi_1(d,d,b)~, \nnb \\
\Pi^{\Xi_b^{\ast -} \rar \Omega_b^{\ast -}     K^{\ast0}} \es 
\sqrt{2} \Pi_1(s,s,b)~, \nnb \\
\Pi^{\Sigma_b^{\ast 0} \rar \Sigma_b^{\ast 0}    \omega} \es 
{1\over \sqrt{2}}\Big[ \Pi_1(u,d,b) + \Pi_1(d,u,b)\Big]~, \nnb \\
\Pi^{\Sigma_b^{\ast +} \rar \Sigma_b^{\ast +}    \omega} \es
\sqrt{2} \Pi_1(u,u,b)~, \nnb \\
\Pi^{\Sigma_b^{\ast -} \rar \Sigma_b^{\ast -}    \omega} \es 
\sqrt{2} \Pi_1(d,d,b)~, \nnb \\
\Pi^{\Xi_b^{\ast 0} \rar \Xi_b^{\ast 0}     \omega } \es
{1\over \sqrt{2}} \Pi_1(u,s,b)~, \nnb \\
\Pi^{\Xi_b^{\ast -} \rar \Xi_b^{\ast -}     \omega } \es
{1\over \sqrt{2}} \Pi_1(d,s,b)~, \nnb \\
\Pi^{\Xi_b^{\ast 0} \rar \Xi_b^{\ast 0}     \phi } \es
\Pi_1(s,u,b)~, \nnb \\
\Pi^{\Xi_b^{\ast -} \rar \Xi_b^{\ast -}     \phi } \es
\Pi_1(s,d,b)~, \nnb \\
\Pi^{\Omega_b^{\ast -} \rar \Omega_b^{\ast -}     \phi } \es
2 \Pi_1(s,s,b)~. \nnb
\eaeeq

Similar relations can be obtained for the heavy baryons containing $c$ quark
by making the replacement $b \rar c$, and adding a positive unit charge to
the charge of each heavy baryon.

\eAPP

\begin{figure}[t]
\begin{center}
\scalebox{0.59}{\includegraphics{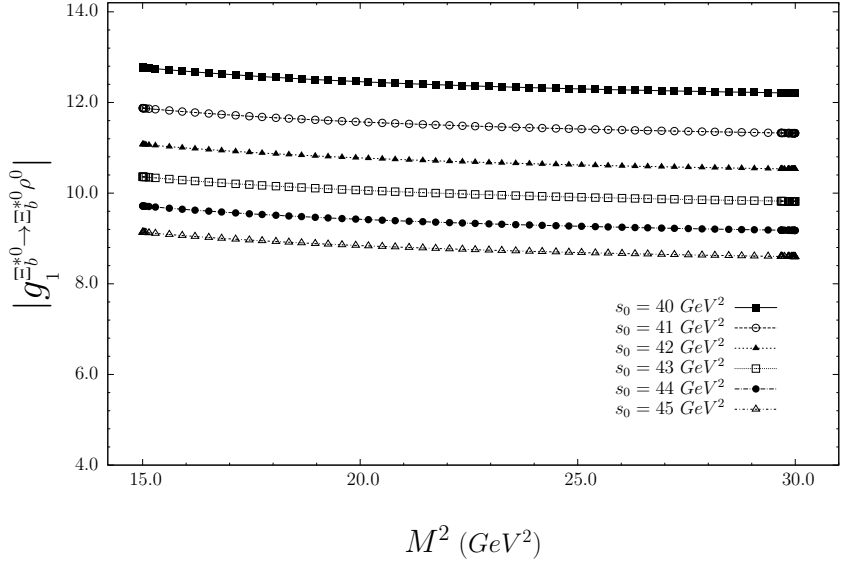}}
\end{center}
\caption{The dependence of the strong coupling constant $g_1$ for
the $\Xi_b^{*0} \rar \Xi_b^{*0} \rho^0$ transition on the Borel mass
parameter $M^2$ at several
different fixed values of the continuum threshold $s_0$.}
\end{figure}

\begin{figure}[b]
\begin{center}
\scalebox{0.59}{\includegraphics{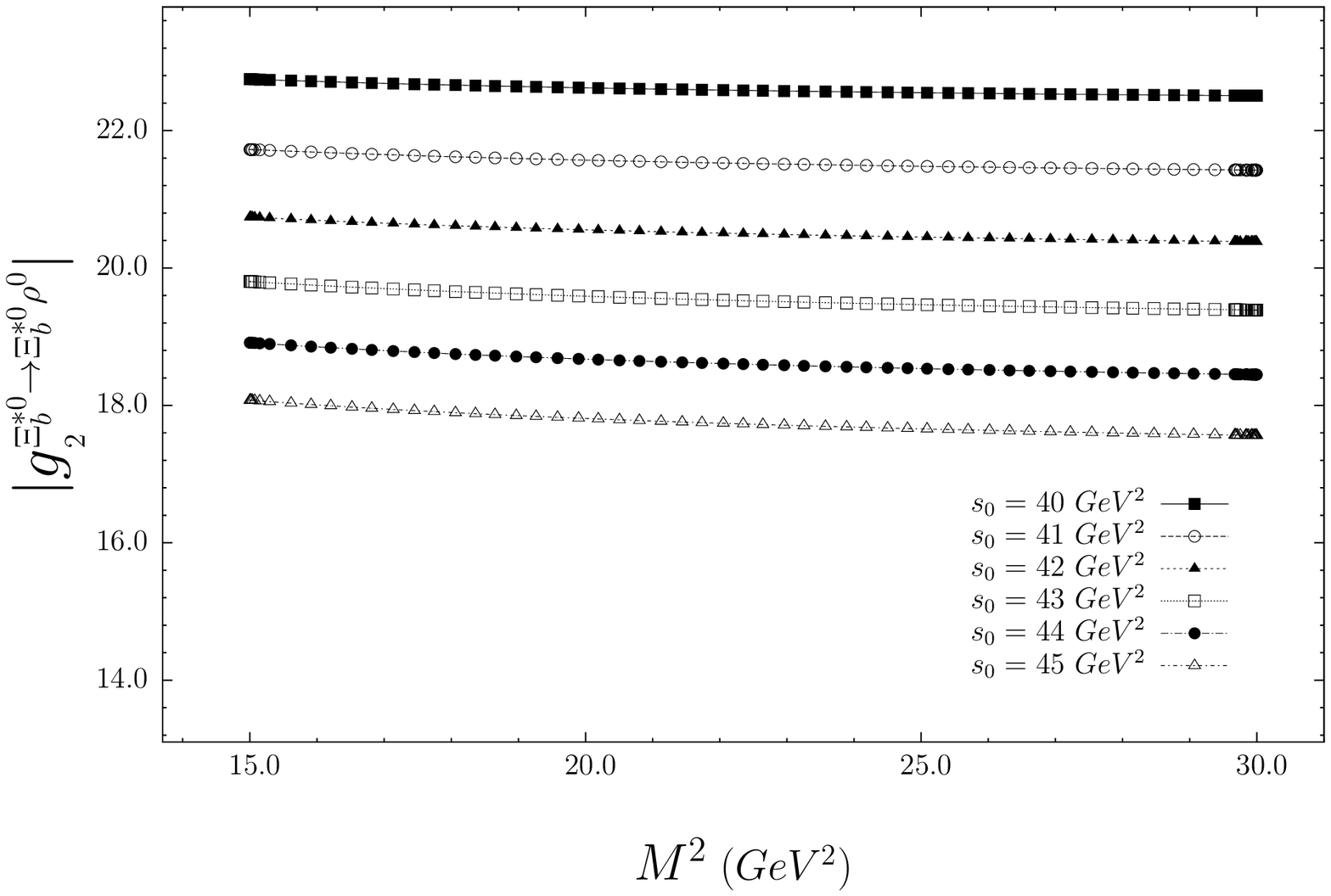}}
\end{center}
\caption{The same as Fig. (1), but for the strong coupling constant
$g_2$.}
\end{figure}

\begin{figure}[t]
\begin{center}
\scalebox{0.59}{\includegraphics{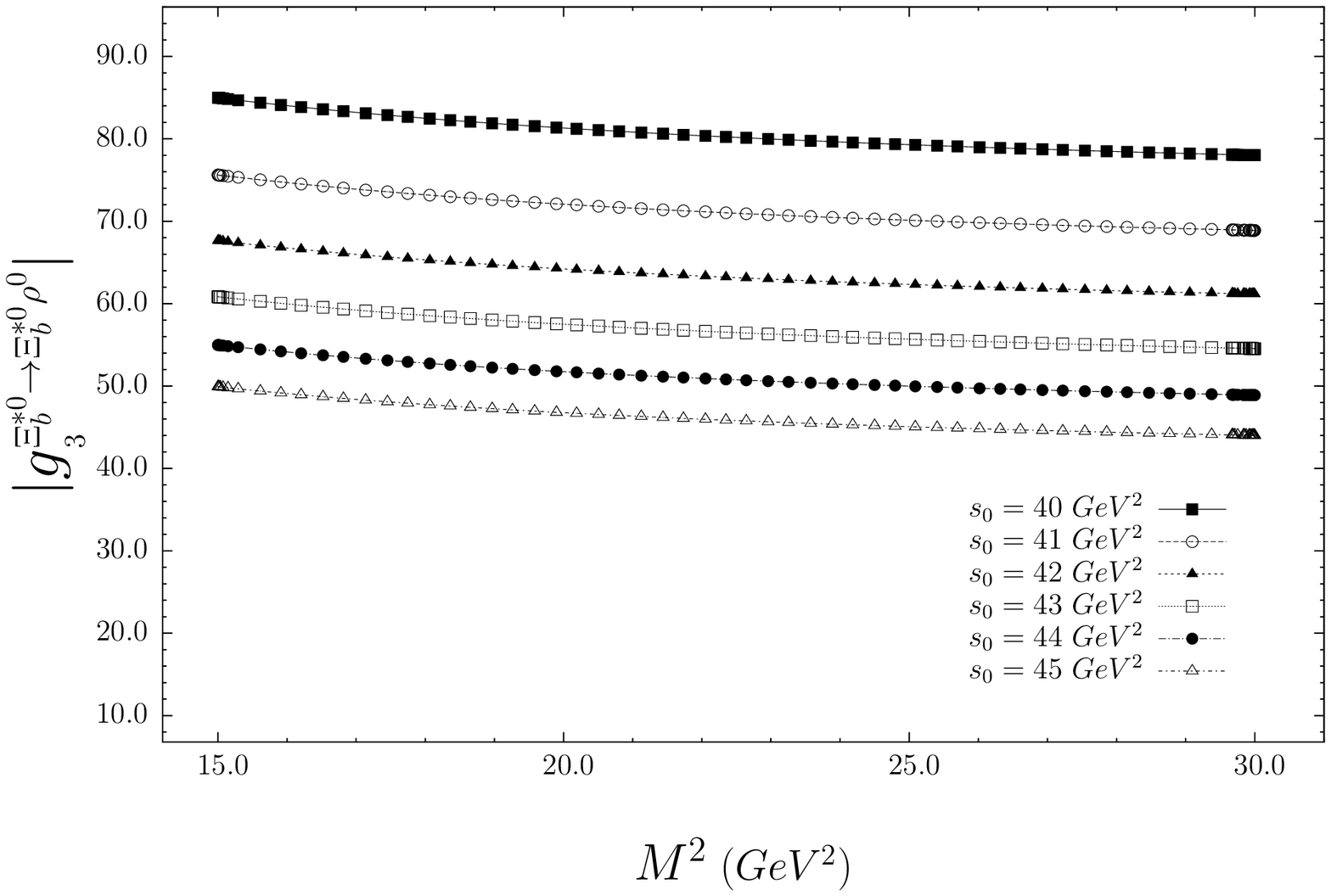}}
\end{center}
\caption{ The same as Fig. (1), but for the strong coupling constant   
$g_3$.}
\end{figure}

\begin{figure}[b]
\begin{center}
\scalebox{0.59}{\includegraphics{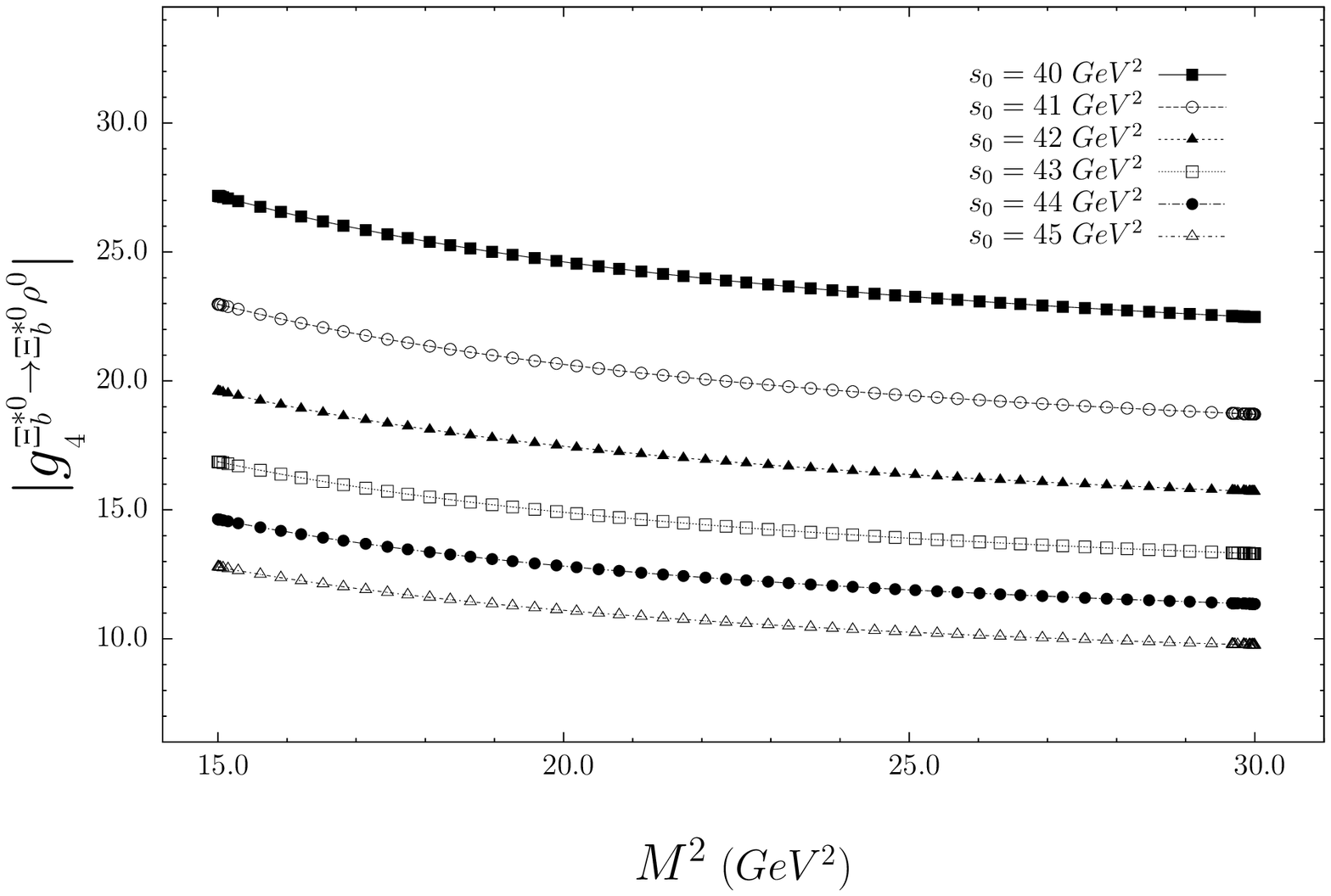}}
\end{center}
\caption{The same as Fig. (1), but for the strong coupling constant   
$g_4$.}
\end{figure}

\end{document}